# Analogue to Digital and Digital to Analogue Converters (ADCs and DACs): A Review Update

*J. Pickering*
Metron Designs Ltd, Norwich, UK

**Abstract**
This is a review paper updated from that presented for CAS 2004. Essentially, since then, commercial components have continued to extend their performance boundaries but the basic building blocks and the techniques for choosing the best device and implementing it in a design have not changed. Analogue to digital and digital to analogue converters are crucial components in the continued drive to replace analogue circuitry with more controllable and less costly digital processing. This paper discusses the technologies available to perform in the likely measurement and control applications that arise within accelerators. It covers much of the terminology and 'specmanship' together with an application-oriented analysis of the realisable performance of the various types. Finally, some hints and warnings on system integration problems are given.

## 1 Introduction

Analogue to Digital and Digital to Analogue Converters (ADCs and DACs) are some of the most important components in measurement and control technology. Their job is to transfer information between the real world and the digital world as faithfully as possible. See Fig. 1 for a typical situation.

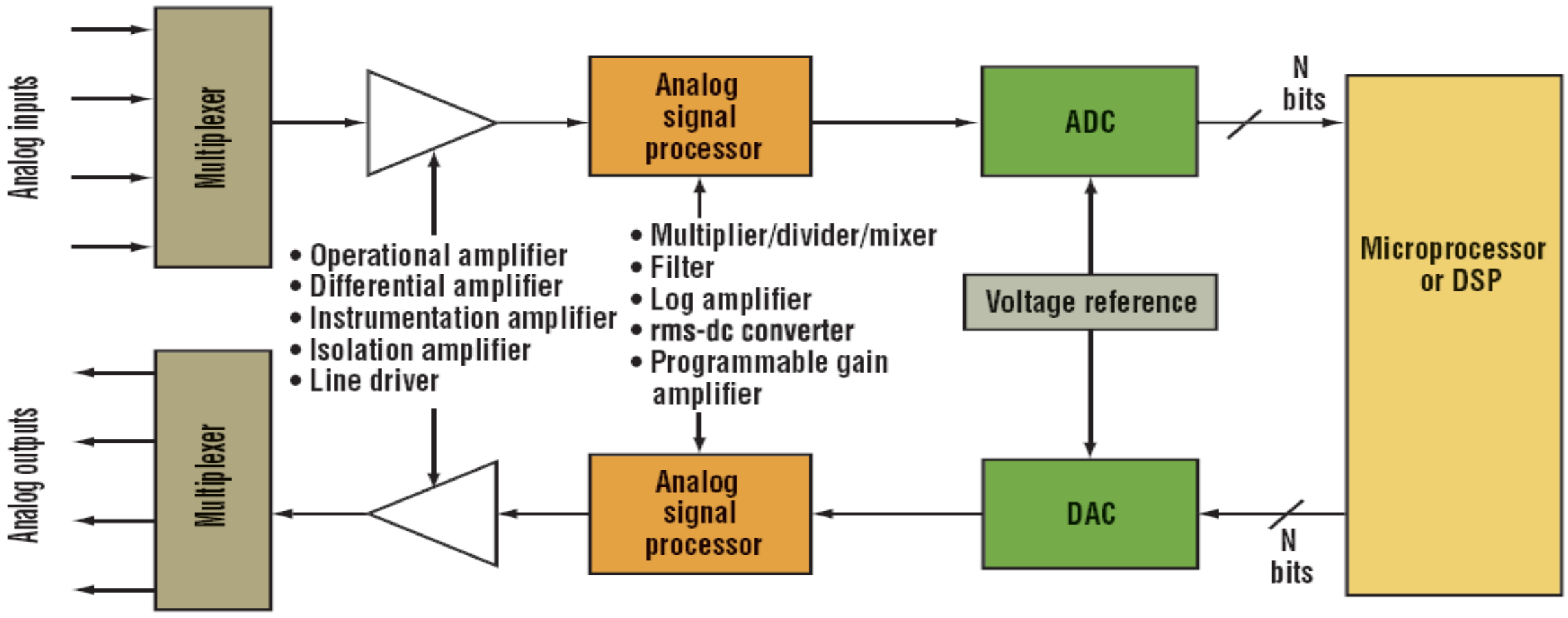

**Fig 1:** Measurement and control loop

Because of the advances in digital processing and its continuing improvement in cost effectiveness it is becoming more and more desirable to trade-off analogue for digital circuitry. An expression for this, commonly used in the communications business, is the trend to push 'digital to the antenna', thus notionally replacing all of the circuitry in a communications receiver with digital processing. The purpose of this article is to consider current commercial ADC and DAC technology suitable for application in measurement and control rather than the very highest performance special-purpose devices. This paper is an update on that appearing at the CERN accelerator school 2004.

## 2 Terms, nomenclature and specifications

Three standards that define specification terms for ADCs and DACs are:

– IEC 60748-4, see Ref. [1];

– IEEE 1241, see Ref. [2];

– Dynamic testing of analogue to digital converters (DYNAD), see Ref. [3].

The IEC standard presently includes a general specification and is currently being updated to include the dynamic specifications that are covered by the other two standards. These, IEEE 1241 and DYNAD (which is the output of a European project SMT4-CT98-2214), concern dynamic specifications mainly defined in the frequency domain. A useful general standard is IEC 60748-4-3, which now covers dynamic performance.

### 2.1 General specifications

The most important terms are best discussed with reference to simplified diagrams showing a 7-bit converter. In all the ensuing discussions we will deal only with converters intended to have equal step sizes (that is, each bit or quantization level is of equal weight). This is not true of certain specialist converters, for example, companding types.

In Fig. 2(a) we have an ideal bipolar converter where the vertical scale is the output code and the horizontal scale is the analogue input. Each step in code represents an analogue increment of $q$, called the step width. As can be seen, starting from 0, each step occurs at ½ $q$ analogue levels. Figure 2(b) shows the same with some random noise added, which results in an uncertainty in the switching thresholds. In these and all of the ensuing descriptions and figures we assume that zero and end point errors have been corrected by a $y = mx + c$ calibration.

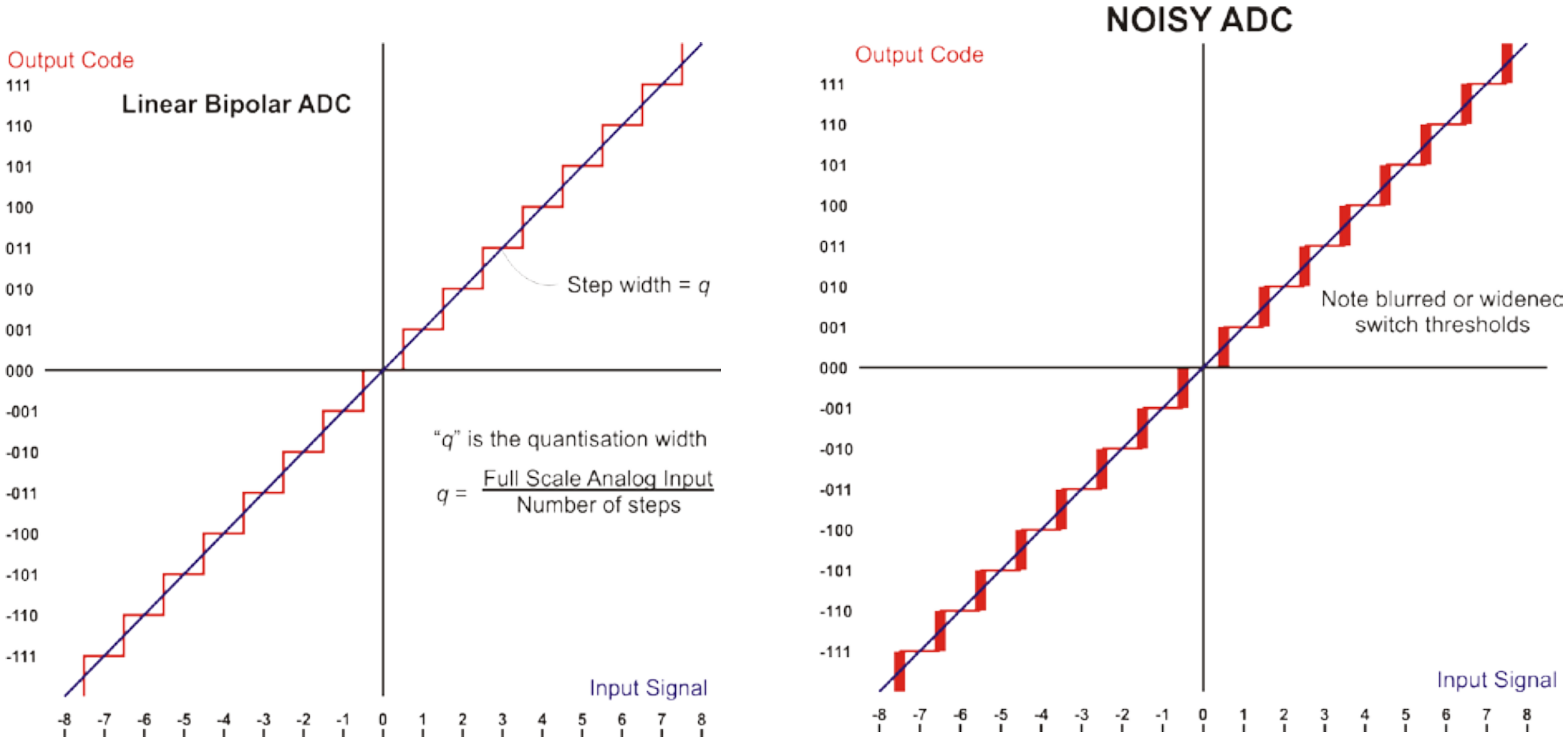

(a) (b)

**Fig 2:** (a) Linear bipolar ADC; (b) noisy ADC

## 2.2 Misunderstood terms: ‘resolution’ and ‘noise’

‘Resolution’, subject to much manufacturer ‘specmanship’, is the smallest discernible increment in analogue terms—however, this should be regarded as ‘repeatably’ discernible since all good measurement should be repeatable, another way of saying that it is not buried in noise or instability. A weakness, in this author’s view, of IEC 60748-4 is that it defines resolution as the ‘nominal value of the step width’, which is far too simplistic and misleading for some types of ADC, resulting in specifications for some being quoted at ‘32 bits resolution’ where reality is nearer 18 bits!

‘Noise’ comes from many sources and is not necessarily random and not necessarily ‘white’ since $1/f$ noise is always present. As far as specification numbers are concerned, noise is also represented very differently for varied applications. In DVMs it usually means ‘peak’ but in most integrated circuit (IC) ADCs it is rms. Presenting it as a percentage of full scale is another variable. DVMs tend to regard full scale as the maximum unipolar excursion possible but IC ADCs regard it as the full range from +ve peak to –ve peak operating range. The result is that the quantitative numbers can vary by 6:1 or more depending on these definitions.

## 2.3 Quantization error or noise $Q_n$

The quantization error of a perfect ADC is that error introduced by the finite number of digital codes, i.e. from the nominal step size. Figure 3 shows this. One can clearly see that the error, as signal is smoothly increased, changes in a saw-tooth manner between $+q/2$ to $-q/2$ where $q$ is the nominal step size.

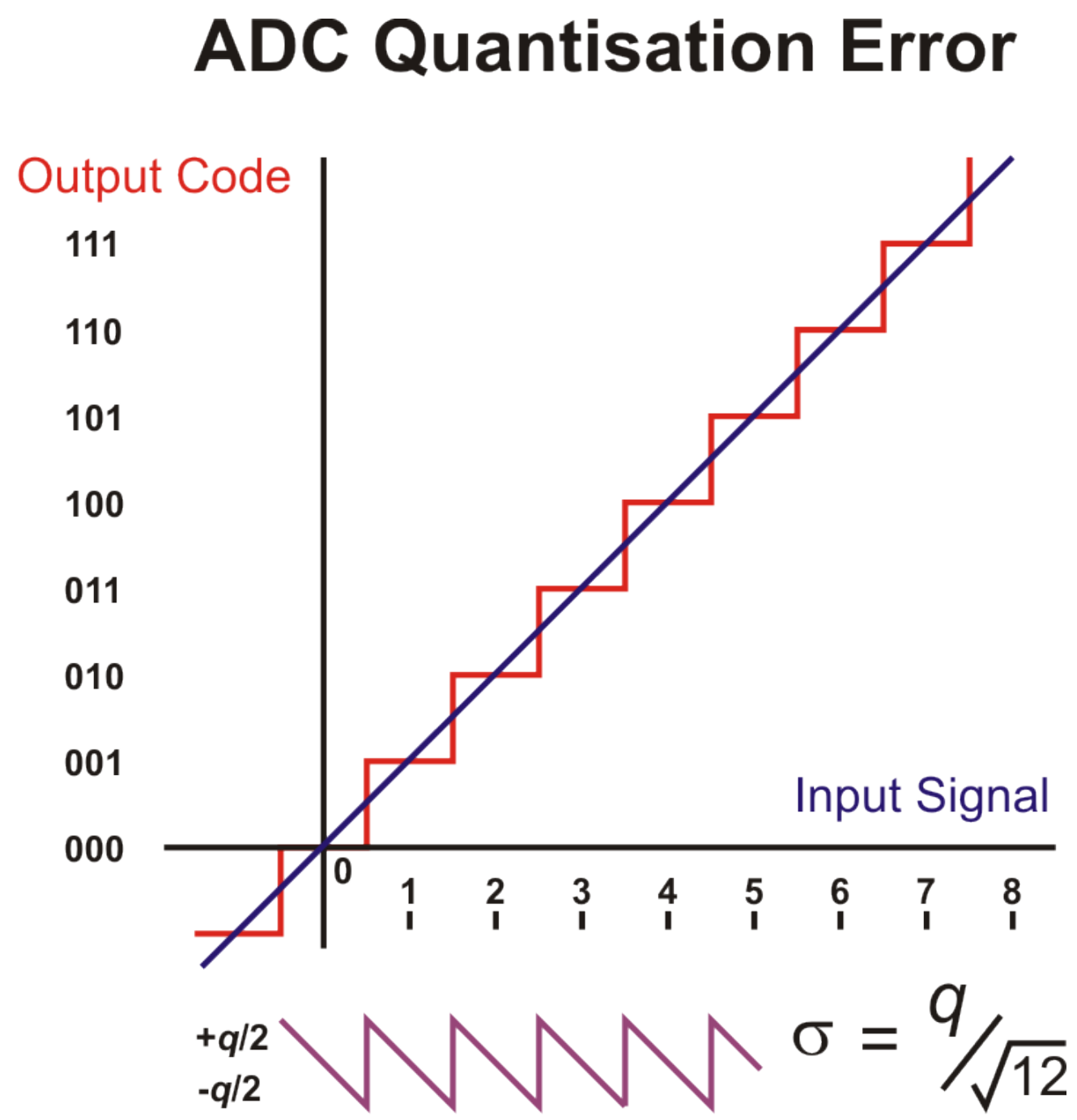

**Fig. 3:** ADC quantization error

This has a standard deviation of $\sigma = q/\sqrt{12}$ and so, in dynamic applications where there is little correlation between the error and the input signal, it is referred to as ‘quantization noise’ and has a magnitude of:

$$Q_n = q/\sqrt{12} \text{ rms.} \tag{1}$$

### 2.4 Differential non-linearity

Differential non-linearity (DNL) is a measure of how individual steps may be in error (for example if comparator bias is incorrect). It is the difference between the ideal step position (in analogue terms) and its actual position. A good ADC will hold this error to $q/2$ (Fig. 4).

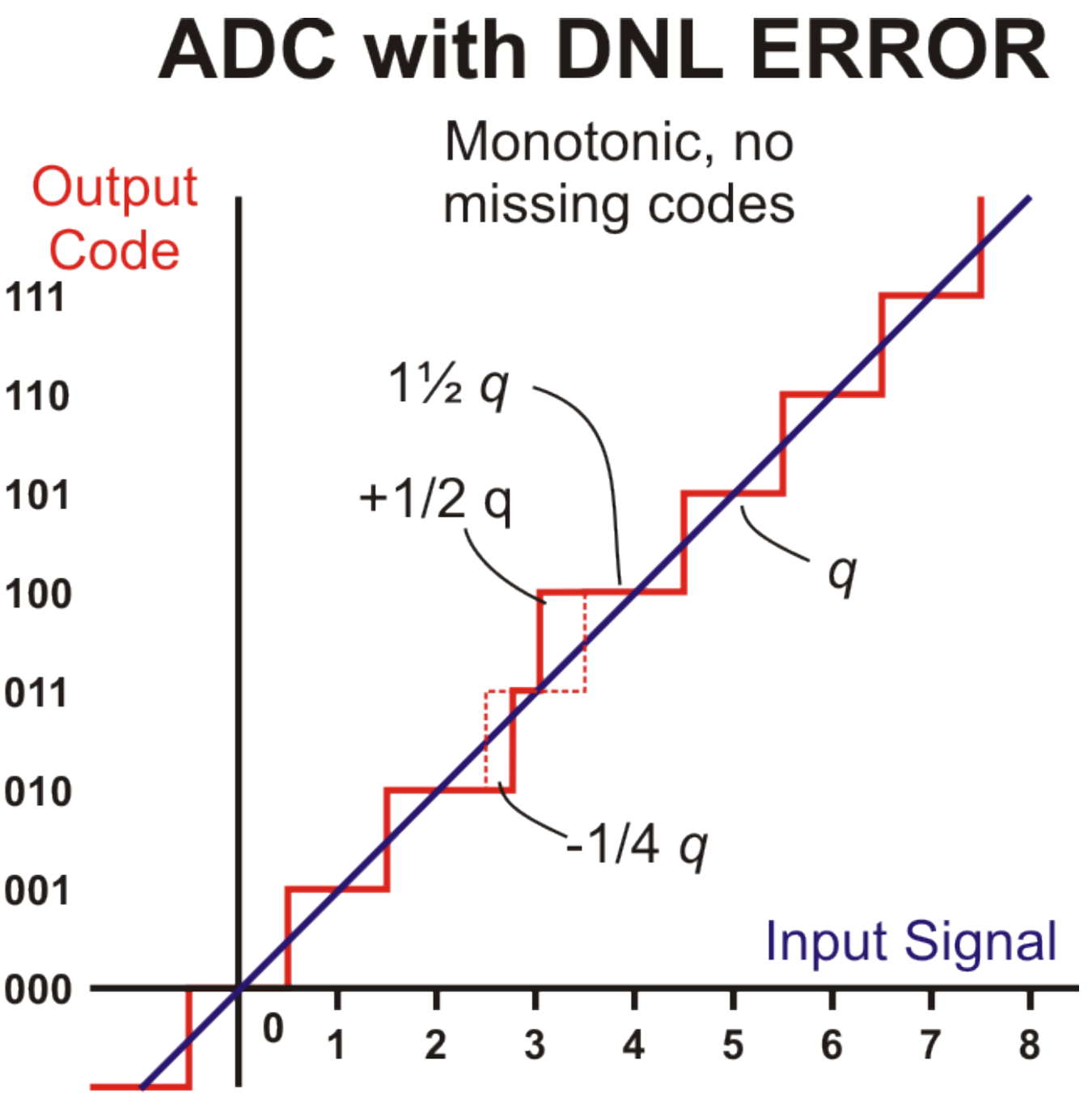

**Fig. 4:** ADC with DNL error. DNL is defined as the difference between ideal and actual step width

### 2.5 Missing codes

If the DNL exceeds $q/2$ it is possible that a missing code can occur, i.e. the converter 'jumps' between non-adjacent codes (Fig. 5).

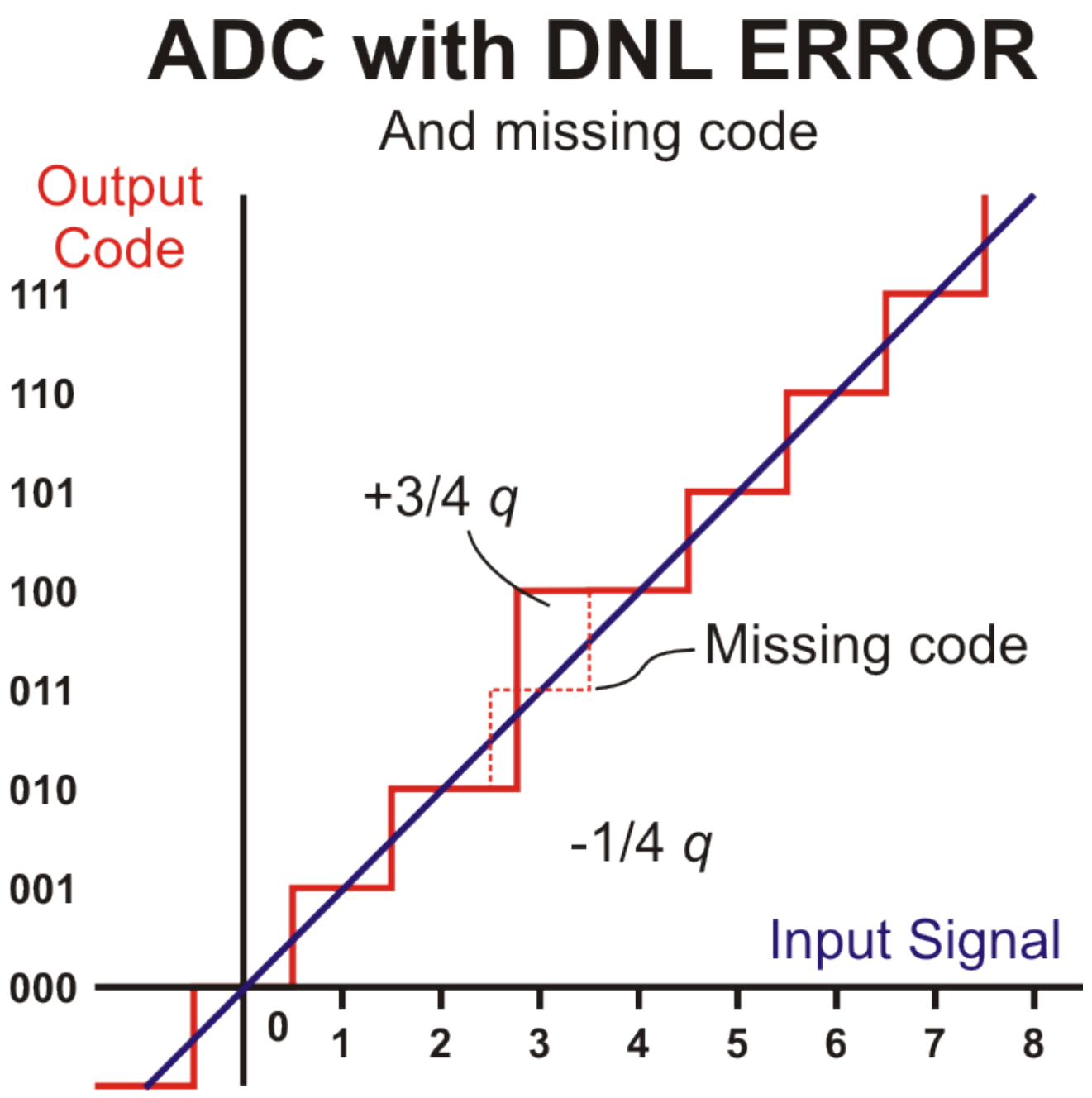

**Fig 5:** ADC with DNL error and missing code

## 2.6 Non-monotonic

Non-monotonic literally means that two different analogue values that are separated by an appreciable increment can produce the same code and this may indeed happen (Fig. 6).

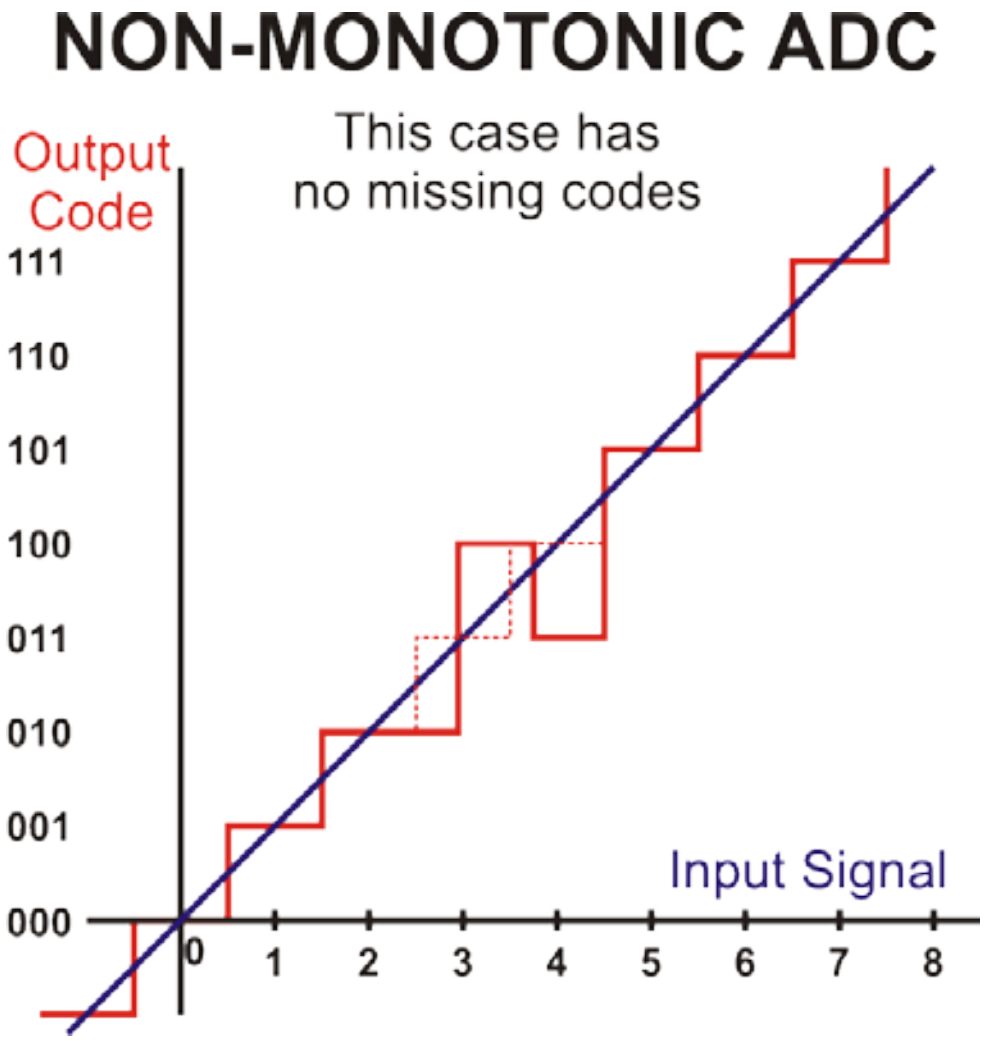

**Fig 6:** Non-monotonic ADC: this case has no missing codes. Non-monotonic is defined as having a negative differential in the output codes.

However, although the term ‘non-monotonic’ is still used it is better defined as a negative differential of code values for smoothly incrementing analogue values. The example is a special case, showing how this can happen without there being a missing code, although a missing code is a more likely result.

## 2.7 Integral non linearity

Integral non linearity (INL) can affect all types of ADC and DAC whereas DNL would not normally affect integrating types such as charge balance and Σ-Δ (delta-sigma). It is defined as the difference between the actual transition at any level and the ideal transition. It occurs where there is an accumulation of very small DNL errors over a range of steps or input signal. It is important to appreciate how it is specified because one can define the ideal transitions as standing on a straight line between zero and full scale endpoints or a ‘best fit’ regression line. In this case INL errors may be distributed symmetrically about a straight line rather than appearing at twice the magnitude, on one side alone (Fig. 7)

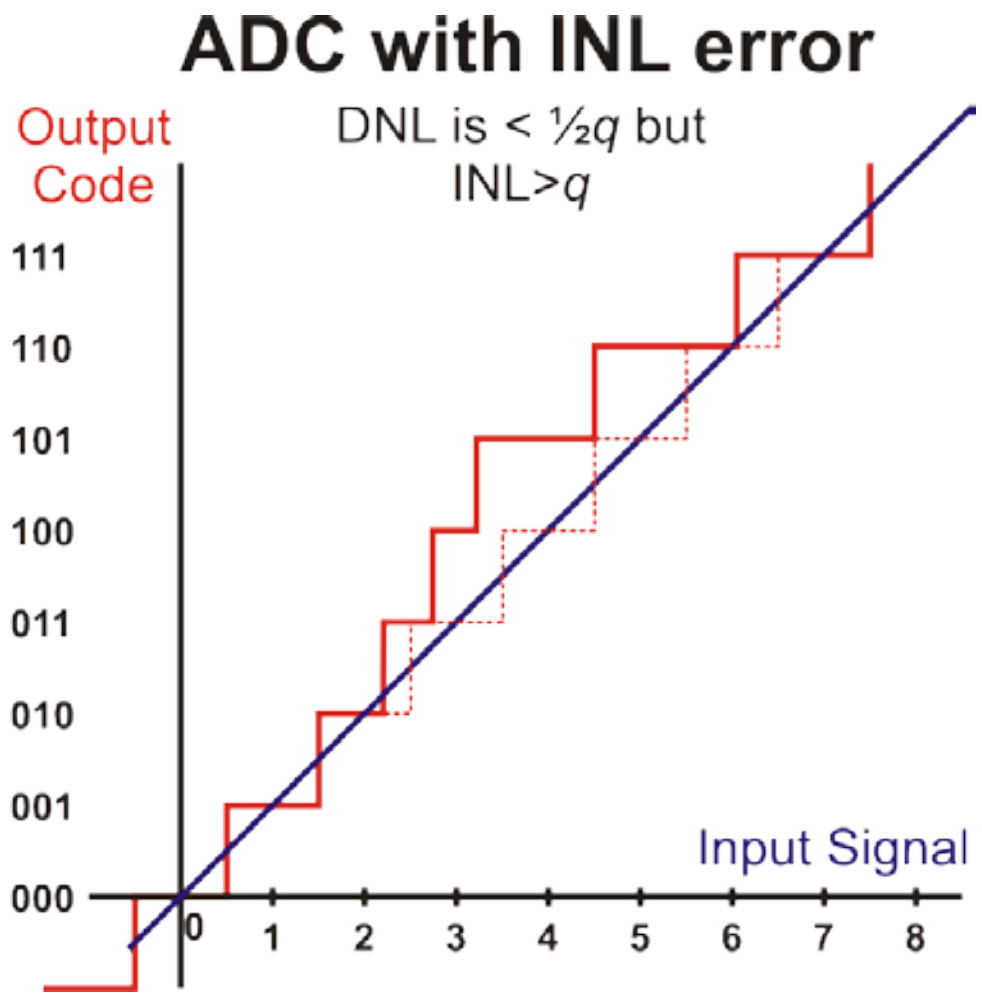

**Fig 7:** ADC with INL error: DNL is <1/2 $q$ but INL > $q$. Defined as extreme deviation from ideal transition

## 2.8 Dynamic terms

ADC data sheets will often be written in terms of AC ‘frequency-domain’ specification and it is important to understand some of the terms used, even if the application is more ‘time-domain’ or DC related.

## 2.9 Signal to noise ratio

Signal to noise ratio (SNR) and the following terms assume that the ADC is to be used to digitize a sine wave set to an amplitude whose peak-to-peak value equals the maximum-to-minimum capability of the ADC. However, this is then translated to ratios of rms values. SNR is the rms ratio of the full scale sine wave to the total noise present, assuming that the quantization noise is random and uncorrelated with signal or other noise. It is a function of the frequency and amplitude of the signal and is therefore specified for a defined signal.

## 2.10 Spurious free dynamic range

Non-linearities and intermodulation products introduce harmonics and spurs into the output spectrum, usually observed through fast Fourier transform (FFT) conversion. The spurious free dynamic range (SFDR) is the range (in dB) between the fundamental and the highest harmonic or spur that occurs within the Nyquist range of ½ the clock rate of the ADC (or over a specified range). Clearly it too will be a function of the amplitude and frequency of the applied signal although the full scale amplitude previously described is usually assumed to give the best ratio.

## 2.11 Signal to noise and distortion

Signal to noise and distortion (SINAD) is commonly used but, whilst useful for an overall system specification, is less so to a designer choosing an ADC to use. It is the SNR but with the spurs and harmonics of the SFDR included and assumed to be an additional uncorrelated noise source. Clearly, when designing a system in practise, one needs to be able to distinguish between truly random terms and signal-dependent terms, especially if the system performs some sort of filtering or averaging that can reduce the effects of random noise but not of distortion.

## 2.12 Effective number of bits

Most IC manufacturers use the term effective number of bit ($N_{ef}$ or ENOB) where DYNAD uses $N_{ef}$ and IEEE 1241 uses ‘effective bits’, E. It is used to give a specification for an ADC’s degradation in resolution when making a measurement where it, the ADC, introduces noise, distortion and spurs. It is thus related to SINAD and can be shown to be (or is defined to be):

$$N_{\mathrm{ef}} = \mathrm{ENOB} = \frac{(\mathrm{SINAD} - 1.76)}{6.02}\,\mathrm{dB}, \qquad (2)$$

where SINAD is expressed in dB of full scale.

## 2.13 Oversampling

Oversampling is used to describe the operation of a converter where samples are taken at a rate higher than two times the input signal’s highest frequency (the Nyquist rate) and some sort of result is obtained by combining these samples. Generally it is the ratio of the sampling rate to the converter’s maximum useable bandwidth. In reality an oversampling ADC is one in which a number of samples of the analogue signal are combined to give a better result (usually higher resolution) than any one sample provides. In order to do this there must be variation in the result for each sample of a perfectly steady signal and this is ensured by inherent or added noise, usually called ‘dither’.

## 3 Types of ADC and DAC

### 3.1 Flash ADCs

Flash ADCs at their simplest can be made with a simple comparator, which senses a voltage threshold and drives a logic level; '1-bit'. More generally they are made with many such arrangements in parallel, each comparator set to a different threshold. The threshold normally divides a full-scale range by the number of comparators. Bandwidths into the GHz are possible (Fig. 8).

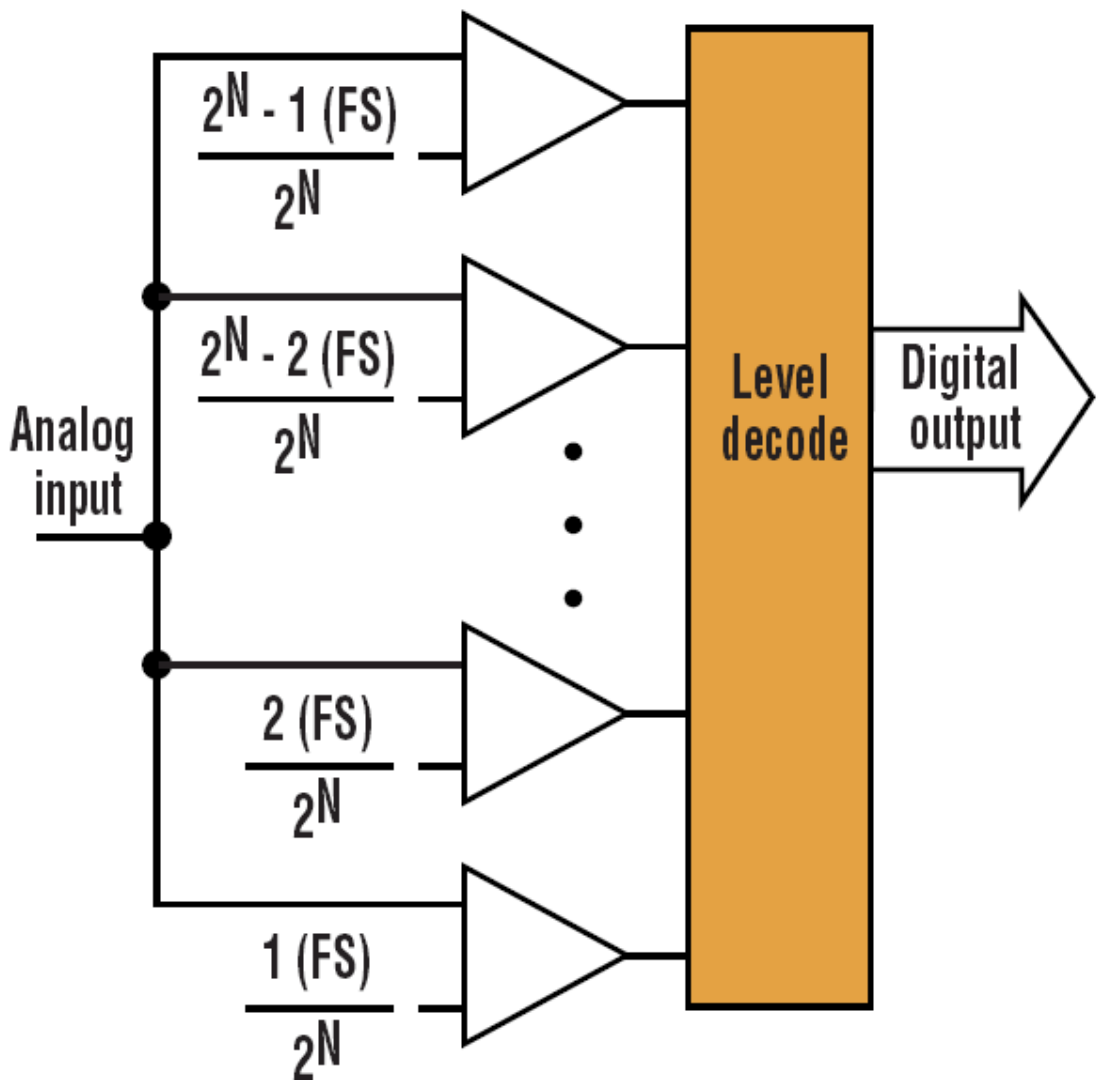

**Fig 8:** Flash ADC architecture. FS, Full-scale analogue input voltage

Modern high-density processing is accomplished mainly in low-voltage technologies that require even greater accuracy and lower noise of comparator thresholds. This, of course, makes the processing and testing expensive.

The Flash ADC is almost unique in not incorporating a DAC in an internal loop. It is this that makes it particularly fast, with a very direct path from analogue input to digital output. It is the workhorse component, at varying resolutions, from 1 bit to 10 bits, within other types of ADC architecture, particularly 'pipeline' and Σ-Δ (where it is often '1 bit').

### 3.2 Pipeline ADCs

Pipeline converters retain most of the bandwidth capabilities of Flash but at higher resolution at the expense of increased sample delay, called latency (Fig. 9).

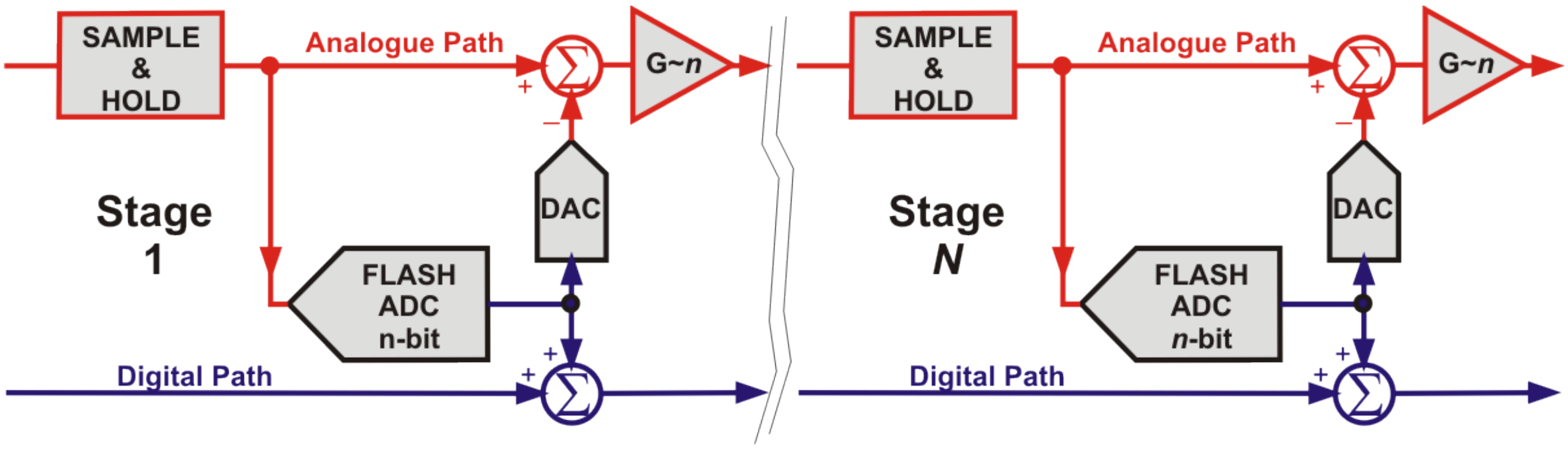

**Fig 9:** Principle of pipeline ADC with $n^N$ bits

They do this by incorporating a number of stages of Flash ADC and DAC in series. This is the pipeline and each stage converts, say, $n$ bits in the Flash but then uses the DAC to subtract the actual Flash output, converted to analogue from the incoming signal, to derive an analogue remainder. This is amplified by $n$ and fed to the next stage to be digitized. Broadly speaking, with n stages, the resolution is $n^N$ but a large degree of self-adjustment has to be incorporated to prevent DNL errors so some resolution is 'wasted' at each stage. It should be evident from this description that the bandwidth can be virtually as high as the Flash converters involved but that there is an addition of delays through the stages—thus leading to considerable latency or group delay.

### 3.3 Successive approximation register ADCs

Successive approximation register (SAR) ADCs have been around for a long time; even the earliest DVMs used them (Figs 10 and 11).

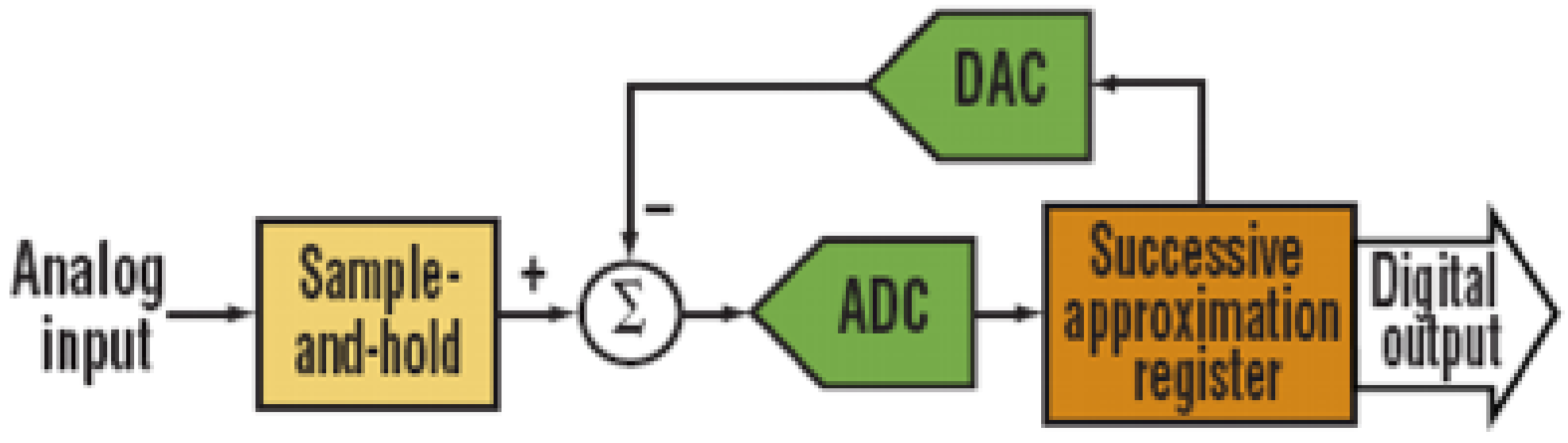

**Fig. 10:** SAR architecture

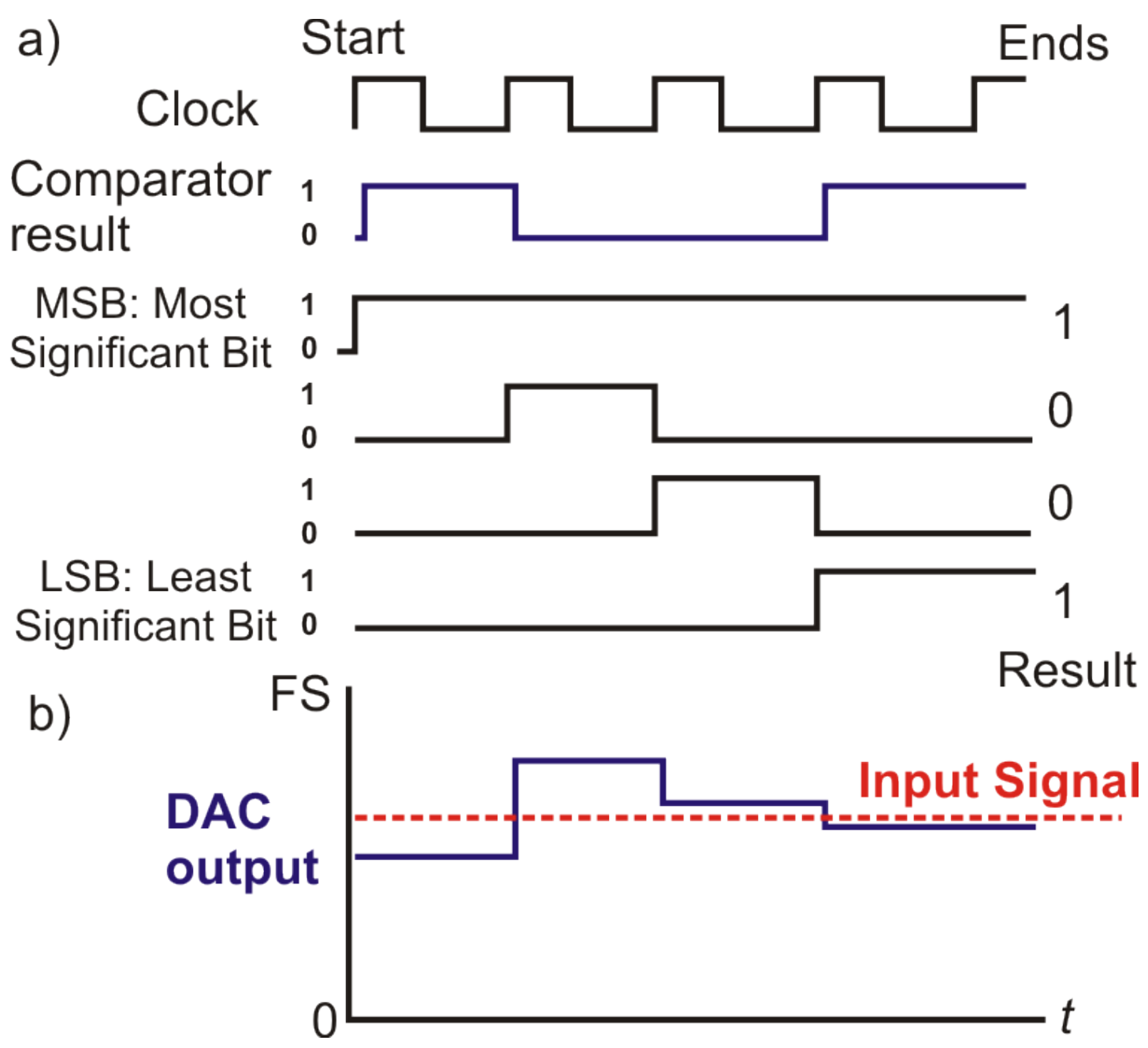

**Fig. 11:** a) SAR logic sequence. b) SAR DAC convergence

The SAR converter operates by testing whether or not the input signal is above or below thresholds set by the ladder DAC. Generally, one starts at ½ of full scale to determine the MSB then sets either ¼ or ¾ scale to determine the next bit and so on. In this example, the ADC in Fig. 10 is simply a comparator, but more complex arrangements can incorporate a Flash converter.

### 3.4 Charge balance, dual slope

The dual slope converter is the most commonly used of the charge balance type. For low-frequency measurement, in simple DVMs and panel meters, it is very cost-effective indeed and well matched to DC measurement with integral noise reduction.

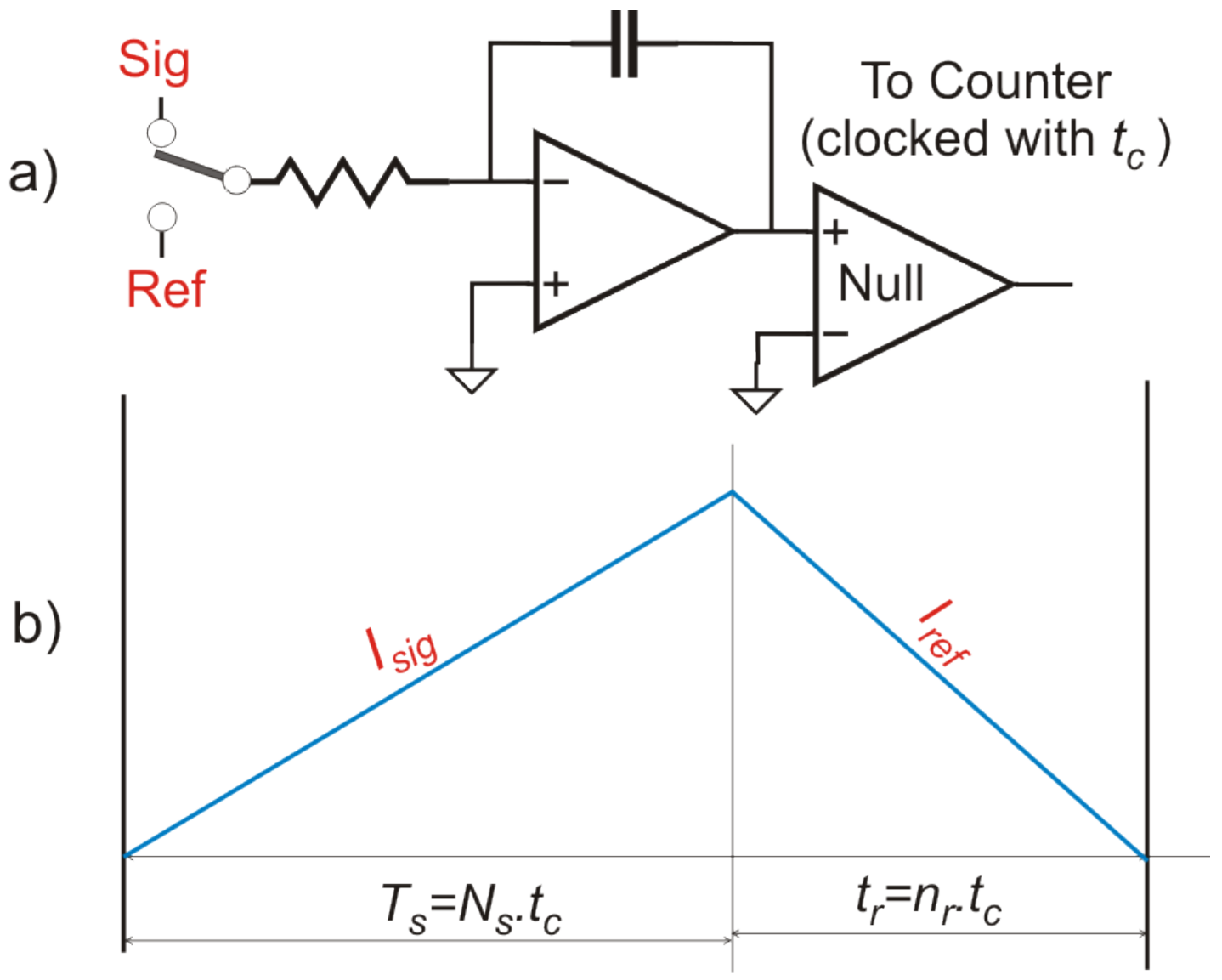

**Fig. 12:** Dual Slope Integrator, a) schematic, b) integration waveform

Basically it utilizes the unknown signal to be measured to charge a capacitance for a pre-set time. It then discharges *through the same components*, with a known reference signal and measures the time this takes. The analogue accuracy is thus dependent only (to a first order) on the reference. Compare this to SAR which also needs as good a reference but in addition a highly critical ladder network of precision resistors or capacitors.

### 3.5 Charge balance, multi-slope

A further refinement that can be used to get something like an order improvement in speed and/or resolution is used in 'top end' DVMs like the Agilent 3458 and Fluke 8508A. Here, in order to improve resolution, reduce the effects of null detector noise and allow smaller integration capacitors to be used, quanta of reference current is removed concurrently with the signal charging. As long as charge balance is maintained and 'accounted for' by the logic any such arrangement is valid.

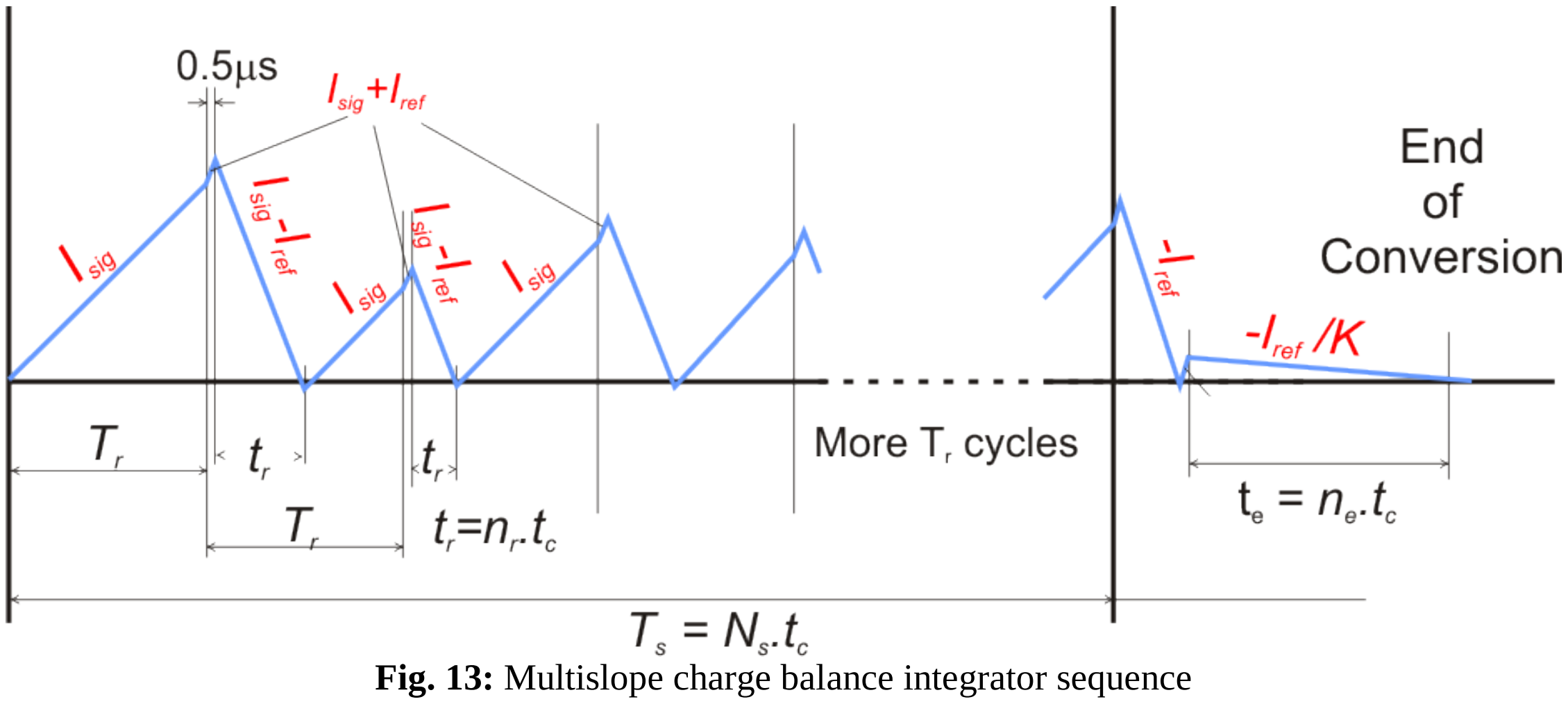

**Fig. 13:** Multislope charge balance integrator sequence

As previously stated there is a large discrepancy in the custom and practise between specifying DVMs and ADCs. Table 1 expresses the specifications of top-end DVMs in a typical ADC format.

**Table 1:** DVM and ADC specification comparison

| Specification | DVM data-sheet specification | ADC 'bit' specification |
|---|---|---|
| Nominal resolution | ±8½ digits | 28+ bits |
| Real (2 σ) resolution | ±7½ digits | 24+ bits |
| Integral non-linearity (INL) | 0.1 ppm ($1 \times 10^{-7}$) | 23 bits |
| Differential non-linearity | No specification. 'Perfect' | 28 bits |

### 3.6 Δ-Σ, delta-sigma, often called Σ-Δ sigma-delta

Delta-sigma (Δ-Σ) is often interchangeably called sigma-delta (Σ-Δ) This conversion technique is usually thought of as a 'charge balance' technique but this is misleading. It is true that over very long periods of time it does maintain a charge balance in its integrator *but* it produces valid, accurate results much faster than would be expected from charge balance equations. In fact, for high resolution, this is several *orders* faster. For example, the CERN Σ-Δ converter produces *independent* 1 ppm resolution conversions in only 1000 clock cycles where dual slope would need 1 000 000. See Ref. [4].

Perhaps the slow take-up of this technology for DC and Low Frequency (LF) metrology has been because this is very difficult to explain with time domain arguments. The frequency domain proponents have no problem! See Refs. [5, 6].

The converter shown in Fig. 14 utilizes a 1-bit DAC (which is very accurate in spite of its low resolution) and the action of the feedback loop is to drive the DAC with a bit stream that balances the incoming signal. Confidence in applying negative feedback-loop theory explains the operation in the time domain!

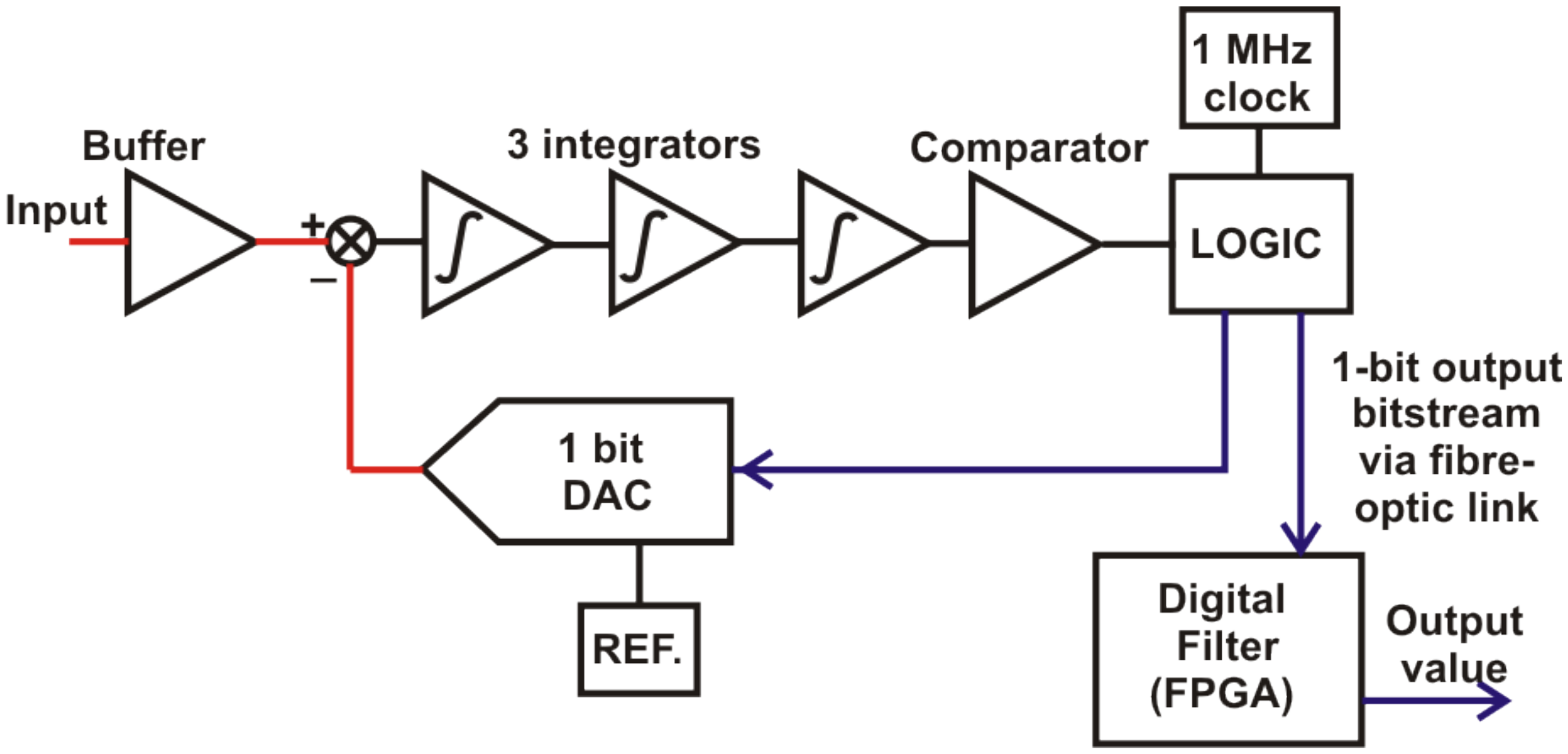

**Fig. 14:** The CERN 1-bit converter

The cumulative gain of the integration stages, forming the modulation filter, is very high indeed, in CERN's case in the order of $4 \times 10^{11}$ at 10 Hz! Clearly, at 10 Hz there can be no significant error at the summing junction.

The cleverness is in achieving this within a loop while maintaining loop stability and, in effect, this is accomplished by feed-forward in the modulation filter, i.e. the integrators. DAC versions of this arrangement, where the input is a bit stream and there is analogue feedback around the loop, are probably the most commonly produced data conversion components of all. They are the basis of the '1-bit' DAC in CD players and have been shipped in millions of units.

Figure 15 shows a simulation result for a 1-bit architecture that uses a digital 4 × 25 stage rolling average filter—it thus obtains all of its information in 100 clocks where dual slope could only achieve 1% resolution. The simulator clearly shows that the resolution far, far exceeds the 'simple' first-order dual slope capability, and further increased digital filtering gives vast improvements in real resolution.

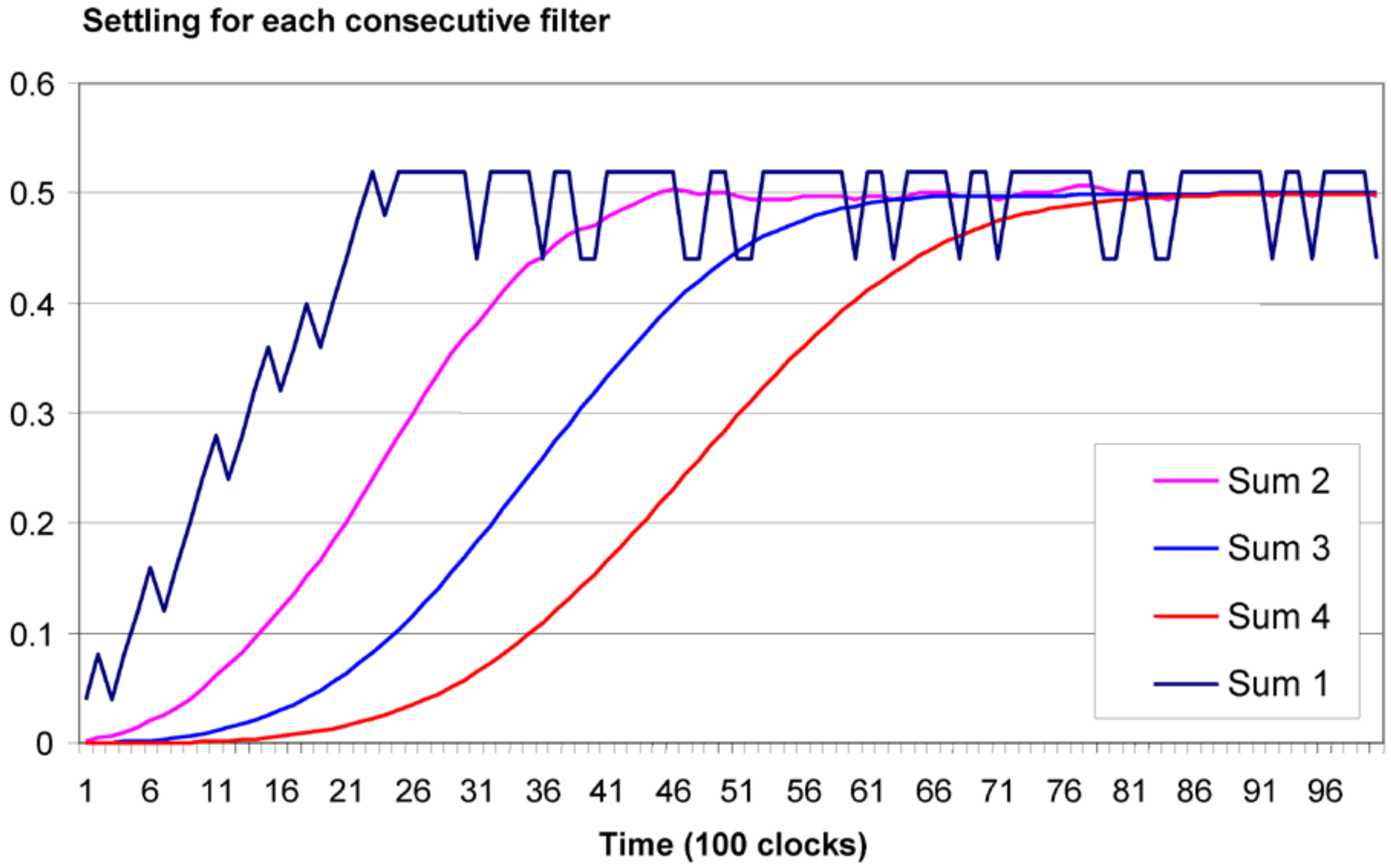

**Fig. 15:** Delta-sigma 1-bit third-order 4 × 25 average filter settling. Note that resolution in 100 clocks is better than 0.1%.

### 3.7 Characteristics of Δ-Σ

Since Δ-Σ seems ideally suited to accelerator DC and LF work, it is useful to look carefully at its characteristics, both good and bad.

Firstly, using IC products from 'merchant semiconductor' companies may give some surprises. Since suppliers are mainly aiming Δ-Σ converters at AC applications, both specification and performance is tailored for this—however the volume of production brings down the cost.

Bandwidths are usually limited but, because of the very high sampling rate, anti-aliasing filters are simple. The multi-stage digital filters allow very high data rates but there are long delays and much of the data is redundant. Clearly, a newly applied signal must replace all data present from the previous one, and this typically takes as many clocks as there are filter stages. However, it is possible to 'look ahead' and take data from early stages of the filter concurrently with the final, more filtered, data. This, of course, can be used for feed-forward in digital control loops.

IC manufacturers promote their devices as, say, '24 bit' or '22 bit' where generally this is the resolution at the 1 σ noise level and with the longest integration times (slowest speed) set, i.e. it is the SNR expressed in bits under the most favourable conditions. Also linearity (INL) is seldom better than 5 ppm (or 18 bits).

There are some problems that are unique to Δ-Σ. Normally, because of the very high loop gain, there is sufficient noise that a sort of chaotic behaviour results and there are no systematic errors in the output due to loop operation. (Of course, there are other systematic errors in, for example, the DAC accuracy.) However, it is well known that certain bit patterns in the feedback loop can be favoured and very low frequency, resonance-like, behaviour results. These conditions, in the frequency domain, look like very low frequency tones and they tend to be common when operation is near zero and the pattern therefore tends to be near symmetric. They are therefore called 'idle tones'. They are, at least in part and perhaps totally, due to unwanted feedback paths (remember how high the loop gain is). They tend to show up more in single chip devices where modulator and digital filters are in the same device.

Of a similar nature and perhaps considered as a 'zero-beat' idle tone, is a characteristic called a 'sticky zero', a hysteresis condition where there is a tendency to lock at zero until the signal is sufficient to overcome the 'glue'—an amount greater than the theoretical resolution. Again, great care in preventing unwanted loops between analogue and digital seems to prevent this.

### 3.8 Choosing the best ADC for the job

Table 2 gives relevant performance against possible requirements. The relative merit scores cannot be taken as correct for all devices under all situations but do at least give an idea. Probably, if one looks at the introduction of new devices from IC manufacturers it would be pipeline and Δ-Σ that have become the most prevalent in recent years.

**Table 2:** ADC type performance comparison

| **ADC type characteristic** | **Flash** | **Pipeline** | **SAR** | **Charge balance** | **Sigma-delta** |
|---|---|---|---|---|---|
| Throughput | Excellent | Very good | Good | Poor | Fair |
| Bandwidth | Excellent | Excellent | Very good | Very poor | Fair |
| Resolution | Poor | Good | Very good | Excellent | Excellent |
| Latency/Hz | Excellent | Fair | Very good | Poor | Fair |
| Linearity/bit | Very good | Good | Fair | Very good | Very good |
| Multiplexing | Excellent | Poor | Very good | Fair | Poor |
| Other | Power! Cost | Very fast clock | DNL stability? | DC only | Easy anti-aliasing |

## 4 Choosing the right specifications

This paper has discussed the relative merits of different architectures and pointed out some specification pitfalls. There remain some aspects of system integration that are not always apparent.

- Many devices have internal references that look to save an external component but beware—internal references are usually of the band-gap type because of the low voltages available. Internally, a band-gap reference is derived from a very low voltage; 60 mV is common and thus is very noisy, particularly with $1/f$ noise. Compensated Zeners are at least an order better than band-gaps.
- The resulting zero performance is then dependent on the reference and internal DAC.
- Ensure that the device you choose produces overload codes and flags that your system can handle—not all are user-friendly.
- Remember that INL is often *orders of magnitude worse* than quoted 'bit specs' and even if the target application system can correct for this there is usually not a specification for stability of INL.

### 4.1 Application problems

Assuming that an application system is designed with care and perhaps verified with simulation there is still likely to be a 'first time round' problem: probably noise. Noise problems are unlikely to show up in simulation and are very difficult to predict—experience suggests the following:

- use ground planes;
- 'bury' HF traces between planes on inner PCB layers;
- if possible make long-path HF signals differential and low voltage level;
- use common mode choke;
- think in terms of current paths—where does HF current flow?
- make signals 'flow' smoothly through the system;
- don't forget that sampling can alias HF noise and bring it down to the frequency of interest;
- HF connects ADC analogue and digital grounds on a plane or with capacitors between planes.

At low frequencies it can also be useful to:

- use 'star points' to control current paths;
- remember that $1/f$ noise cannot be averaged out totally and becomes a limit to achievable performance;
- use chopper stabilization to overcome drifts and $1/f$ noise in amplifiers. With the availability of suitable components it is now cheap and simple to do so;
- prevent differential temperatures from being developed across sensitive circuit areas, i.e. prevent heat flow.

### 4.2 An application example

A very high performance ADC has been developed, see Fig. 16.

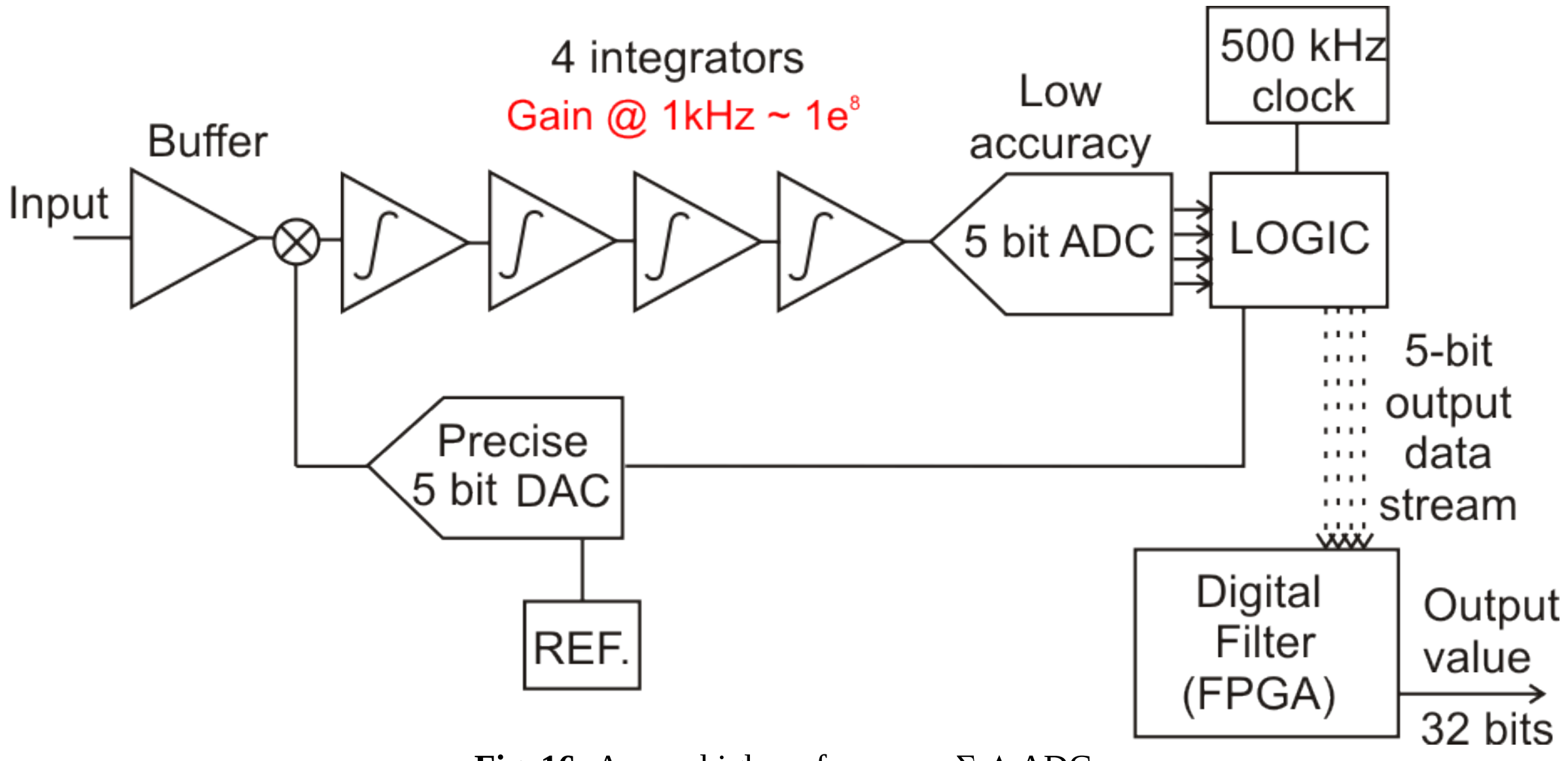

**Fig. 16:** A very high performance Σ-Δ ADC

In this example the 1-bit DAC is replaced with a 5-bit pulse width modulation (PWM) DAC and the output comparator replaced with a 10-bit pipeline ADC of which 5 bits are used. This architecture can achieve '28-bit' performance in resolution and 24 bits in linearity.

In order to achieve 5-bit resolution of the PWM at up to 500 KHz it is necessary to clock the pipeline ADC at 20 MHz. We thus have the difficult combination of 20 MHz clocking on the same PCB near to where 100nV DC performance is needed!

Figure 17 shows the PCB layout arrangement chosen, about half of the full size.

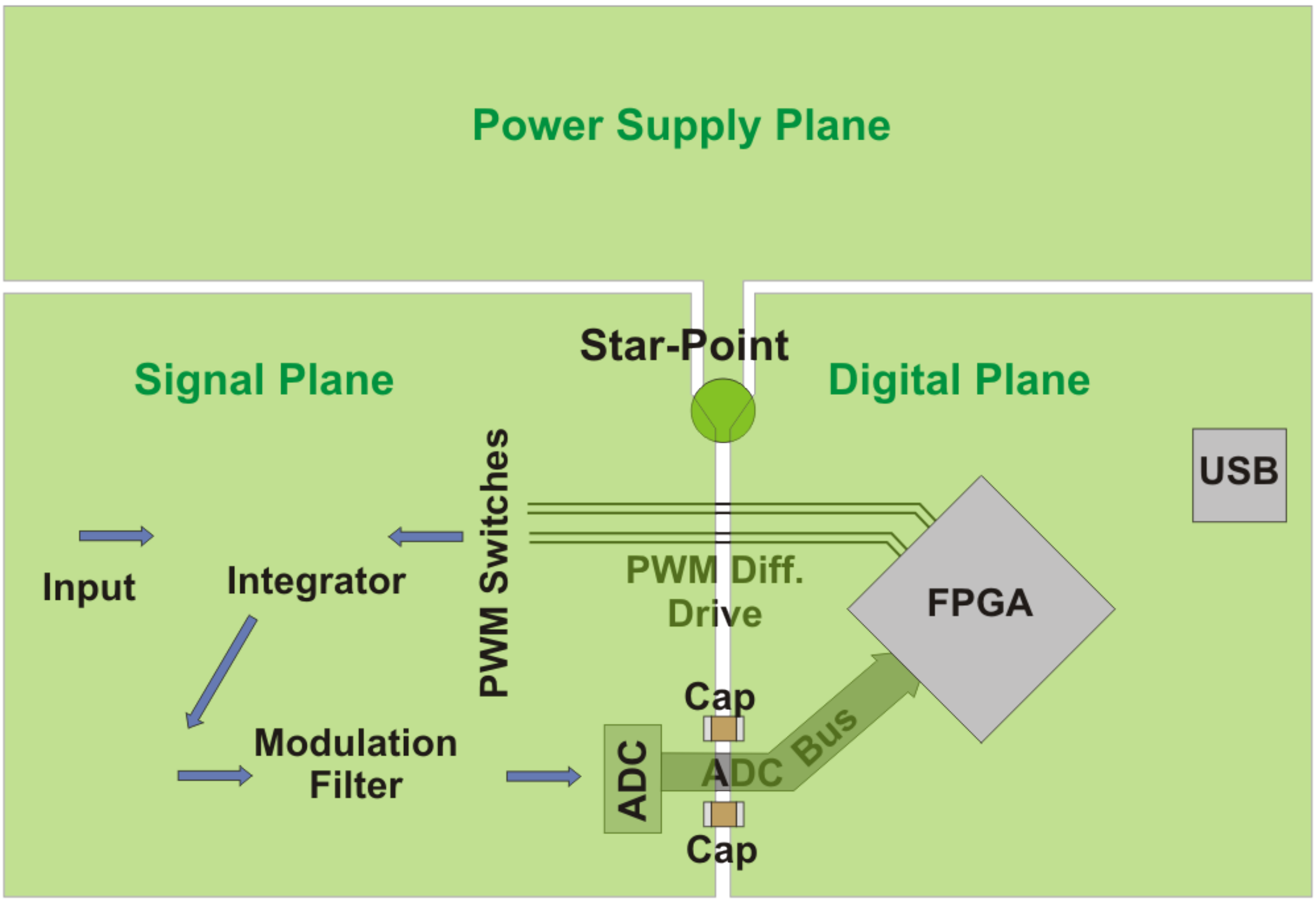

**Fig. 17:** Practical board layout

The PWM drive (with sub-ns edges) is passed from the field-programmable gate array (FPGA) to the analogue section of the circuit (the signal plane in Fig. 17) with differential signal traces (and indeed, the analogue switching is differential). A 'star-point' localizes LF currents and prevents those in power supply and digital planes from causing voltage drops in the signal plane.

This arrangement can be considered as a good example of any high-performance ADC application. Currents are controlled or 'steered' by the use of different ground planes, and circuitry is arranged for even current flow. In fact, the above PCB had a problem because signal-related ground currents could feedback from the output of the modulation filter to the PWM switches and integrator, causing idle tones and sticky zeros. The solution was to split the ground plane between these areas with a slot in the PCB. Furthermore, the sampling ADC drives very fast transient currents into the input capacitance of the FPGA and the two caps are fitted to 'encourage' local HF current loops rather than letting the return current pass back through the 'star-point'. Always remember to think about where the currents flow whether they are LF signal-related or HF with the ability to cause interference noise. A later version included analogue isolation between the modulation filter and the sampling ADC with the ADC placed on the digital plane.

## 5 Finally, the future—cryogenics?

The steering magnets at CERN and in many other accelerator projects are cryogenic. Much fundamental metrology is now based on quantum physics operating in liquid helium. Why not put some of the measurement in the cryogenic environment? Figure 18 is a suggestion and has been in operation at the UK's National Physical Laboratory.

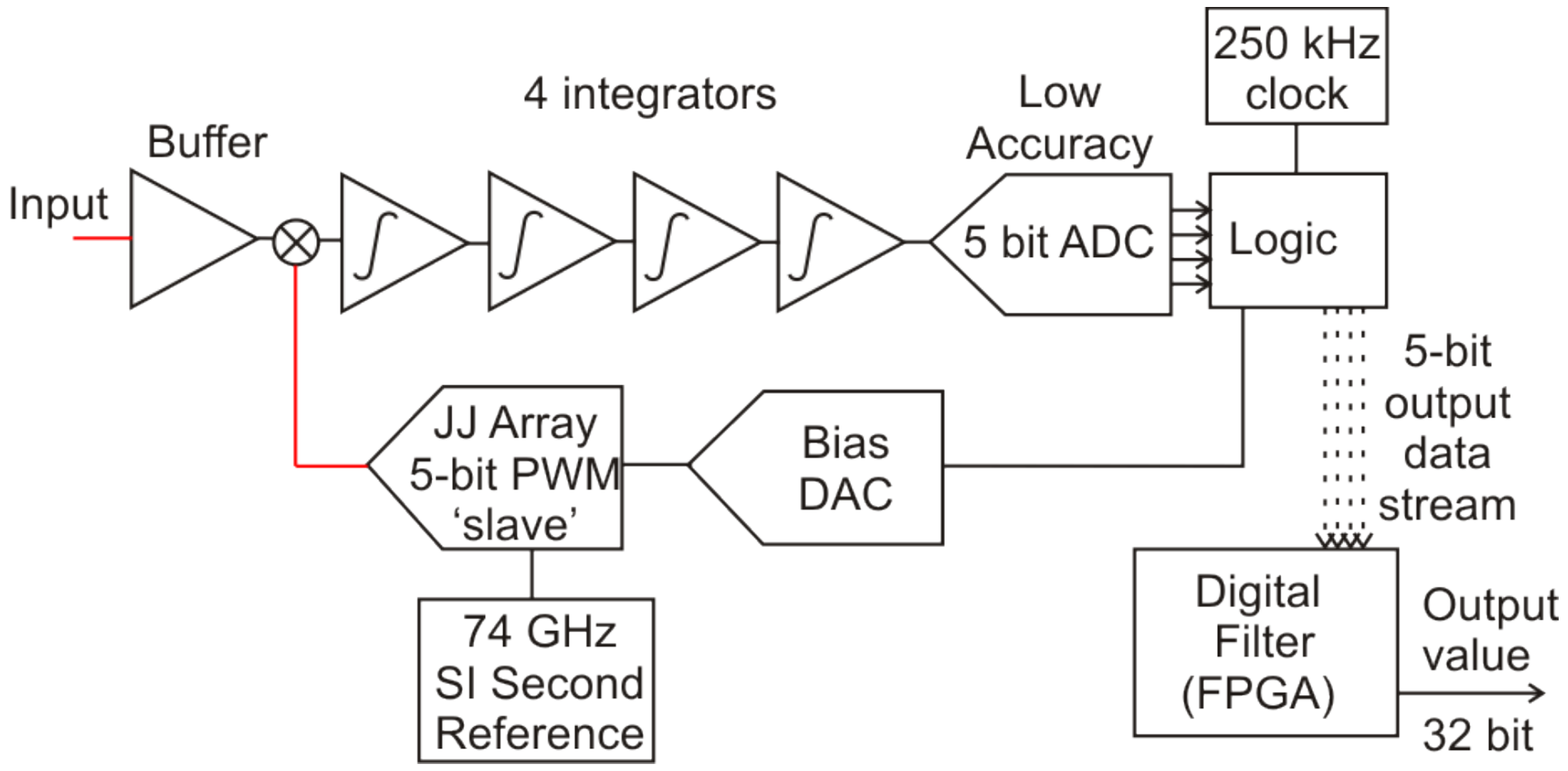

**Fig. 18:** Integration of Σ-Δ modulator with Josephson junction array (JJ)

Its operation is quite simple—the bias DAC is a two-level PWM generator that turns ON and OFF precisely the bias to the Josephson Junction (JJ) array. The operation of the JJ is to lock the amplitude of the resulting pulse to a quantum level dependent only on fundamental constants and on the SI second. See Refs. [7, 8]. Sadly, the speed with which the array could be switched was limited by the capacitance and delays in the JJ drive electronics but with further development could become a 'quantum voltmeter'.

## References

[1] Semiconductor devices – Integrated circuits – Part 4-3: Interface integrated circuits – Dynamic criteria for analogue-digital converters (ADC), IEC 60748-4-3 ed1.0 (2006). http://webstore.iec.ch/webstore/webstore.nsf/artnum/036677!opendocument

[2] 1241-2000 – IEEE standard for terminology and test methods for analog-to-digital converters (2000). http://standards.ieee.org/findstds/standard/1241-2000.html

[3] C. Morandi *et al*., *Comp. Stand. Inter.* **22** (2000) 113. http://www.sciencedirect.com/science/article/pii/S0920548900000374

[4] J.G. Pett, A high accuracy 22 bit sigma-delta converter for digital regulation of superconducting magnet currents, Advanced A/D and D/A Conversion Techniques and their Applications, IEE Conf. Pub. 466 (1999), p. 46. IET, Stevenage UK.

[5] J. Candy and G. Temes, *Oversampling Delta-Sigma Converters* (IEEE Press New York, 1992).

[6] R. Schrier and G. Temes, *Understanding Delta-Sigma Data Converters* (IEEE Press, Wiley Interscience, 2005, NJ, USA). ISBN: 978-0-471-46585-0

[7] P. Kleinschmidt, P. Patel, J.M. Williams, T.J.B.M. Janssen, R. Behr, J. Kohlmann, J. Niemeyer, J. Hassel and H. Seppa, *Arbitrary waveform synthesis using binary Josephson junction arrays*, 10th British Electromagnetic Conf. Conference Digest, Harrogate, 2001.

[8] C.A. Hamilton, C.J. Burroughs and R.L. Kautz, *IEEE Trans. Instrum. Meas*. **44** (1995) 223.

There is also much valuable material on many semiconductor manufacturers' web sites.